\documentclass{aa}  

\usepackage{amsmath}
\usepackage{amssymb}
\usepackage{graphicx}
\usepackage{dcolumn}
\usepackage{bm}
\usepackage[switch,mathlines]{lineno}
\usepackage{upgreek}

\usepackage{xcolor}

\usepackage{lipsum}
\usepackage{float}
\usepackage{graphicx}
\usepackage{txfonts}

\usepackage{aas_macros}

\title{Dynamics of energetic electrons scattered in the solar wind}
\subtitle{Magnetohydrodynamics and test-particle simulations}

\author{A. Houeibib \inst{1}
  \and F. Pantellini \inst{1}
  \and L. Griton \inst{1}}

\begin{document} 

    \institute{LESIA, Observatoire de Paris, Université PSL, CNRS, Sorbonne Université, Université Paris Cité, 5 place Jules Janssen, 92195 Meudon, France}

   \date{Accepted December 29, 2024}

    \abstract{
    We model the transport of solar energetic particles (SEPs) in the solar wind. We propagated relativistic test particles in the field of a steady three-dimensional magnetohydrodynamic simulation of the solar wind. We used the code MPI-AMRVAC for the wind simulations and integrated the relativistic guiding center equations using a new third-order-accurate predictor-corrector time-integration scheme. Turbulence-induced scattering of the particle trajectories in velocity space was taken into account through the inclusion of a constant field-aligned scattering mean free path $\lambda_\parallel$. 
    We considered mid-range SEP electrons of $81\:{\rm keV}$ injected into the solar wind at a heliocentric distance of 0.28 AU and a magnetic latitude of $24^{\circ}$. For $\lambda_\parallel =0.5\:{\rm AU}$, the simulated velocity pitch angle distributions agree qualitatively well with in situ measurements at 1 AU. More generally, for $\lambda_\parallel$ in the range 0.1 to 1\:AU, an energy-loss rate associated with the velocity  drift of about $10\%$ per day is observed. The energy loss is attributable to the magnetic curvature and gradient-induced poleward drifts of the electrons against the dominant component of the electric field. In our case study, which is representative of the average solar wind conditions, the observed drift-induced energy-loss rate is fastest near a heliocentric distance of 1.2 AU. We emphasize that adiabatic cooling is the dominant mechanism during the first 1.5 hours of propagation. Only at later times does the drift-associated loss rate become dominant.}


   \maketitle
%

\section{Introduction}

Solar energetic particles (SEPs) have attracted attention since \cite{Forbush_1946} first imputed the sporadic 
increase in the cosmic-ray flux measured in ionization chambers to GeV protons ejected by the Sun during flares (see \cite{Cliver_2008} for more details). 
In the same epoch, radio astronomers started to detect flare-associated radio bursts in the frequency 
range < 100MHz \citep{Payne-Scott_1947}. Some of these, in particular, the type II and  type III bursts (see \cite{Wild_etal_1950}), were later suggested to be 
produced by a plasma instability excited by a stream of fast particles \citep{Ginzburg_1958}. 
By the beginning of the 1960s (see \cite{Wild_etal_1963}), it was generally accepted that radio bursts are due to the interaction of energetic particles with interplanetary plasma. These particles are accelerated either impulsively at the sites of magnetic reconnection in the solar corona or gradually, at the shock ahead of a coronal mass ejection (CME) \citep{Reames_1999,Desai_etal_2016,Klein_etal_2017}.

After it was established that solar flares and CMEs can accelerate ions and electrons to relativistic energies, the question of their propagation throughout the turbulent heliospheric plasma arose. To model this propagation, two approaches were attempted so far.  The first approach is to consider that the particle diffusion is a Markovian process that is conveniently described by the Fokker-Planck equation, which describes the propagation and the diffusion in phase-space 
of an initial probability distribution of particles \citep{Parker_1965}. 
The linearity of the Fokker-Planck equation enables analytic solutions  in some simple cases, such as the case of the diffusion of energetic particles in a constant and  nonmagnetized solar wind (see \cite{Parker_1965}). However,  for more realistic wind models, the  Fokker-Planck equation has to be solved numerically on a five-dimensional grid spanning three spatial and two velocity dimensions in phase space, which can be computationally very 
costly (e.g., \cite{Wijsen_2020,Wijsen_2022}). The computational costs can be reduced by integrating the  trajectory
of N test particles on a three-dimensional spatial grid with prescribed electric and magnetic fields. 
The second approach consists of computing the trajectory of a large number of individual particles in a prescribed electromagnetic field.
Scattering is then modeled by hard-sphere-type collisions given a prescribed collisional mean-free path $\lambda$ along the magnetic field of the particle guiding center and a post-collision angular distribution law. We follow the second approach here. Our treatment is similar to but different from the one adopted by  \cite{Marsh_etal_2013,Dalla_etal_2013,Dalla_etal_2015}, where the scattering events were Poisson distributed in time with an average scattering time $\lambda/v$, that is, the average time interval between successive scattering events is constant instead of the distance of the scattering centers along the field line. The underlying physical model differs from the classical quasi-linear Fokker-Planck model, which rests on the assumption that particles change their trajectory through a succession of small-angle deviations and on the assumption of quasi-isotropy of the velocity distribution function \citep[see e.g.][]{Jokipii_1966,Jokipii_1968,Hasselmann_1968,Schlickeiser_1989}. 
The differences can be reduced by increasing the number of test particles N and decreasing the Knudsen number $\lambda/L_{\rm B}$ ($L_{\rm B}$ is the spatial scale of the variation in the magnetic field) and provided that the adopted post-collision angular distribution law drives the velocity distribution toward isotropy (see Sect. \ref{sec:pitch}).  

We considered the propagation of electrons as test particles in the electromagnetic field of a steady-state solar wind from a 3D magnetohydrodynamic (MHD) simulation, which conveniently describes the large scales. We considered solar energetic electrons of 81 keV, which are in the mid-range for SEP electrons (see Fig 1 in \cite{droge_2018}). 
For an 81 keV electron in a magnetic field $B = 1 \:{\rm nT}$ , the Larmor radius is $r_L = 678\:{\rm km}$ ($2400\:{\rm km}$ for 1 MeV electrons). This is much smaller than the order AU scales of the MHD simulation, so that we consistently integrated the equations of motion in the relativistic guiding-center approximation (GCA). 
The structure of the GCA equations is far more complex than the full equations of motion, but it has the considerable advantage of removing the constraint of an integration time step that is smaller than the particle gyration period, which allowed us to follow the particles with higher precision over longer time periods and at substantially  lower computational costs. 
We only considered scattering-induced parallel diffusion. We did not consider perpendicular diffusion, which for these energies is known to be smaller by a factor 10 to 100 near a heliocentric distance of 1 AU \citep{Palmer_1982,Chhiber_etal_2017}.

\section{\label{sec:Model}Model and numerical setup}
\subsection{\label{subsec:MHD-part}3D MHD simulation of the solar wind }
To simulate the solar wind, the adiabatic ideal MHD equations were integrated numerically using the TVDLF scheme and the Woodward slope limiter of the code MPI-AMRVAC. The code has been widely used to simulate solar and astrophysical flows \citep[see e.g.][]{Xia_etal_2018}. The simulation domain was a Sun-centered spherical grid of size $144 \times 48 \times 128$ in $[r,\theta,\varphi]$, where $r$ is the radial distance, extending from $r=0.139\:{\rm AU}$ to $r=13.95\:{\rm AU}$, $\theta$ is the polar angle, ranging from $0$ to $\pi$, and $\varphi$ is the azimuthal angle, spanning from $0$ to $2 \pi$.
\begin{figure}[h]
\centering
    \includegraphics[width=\linewidth]{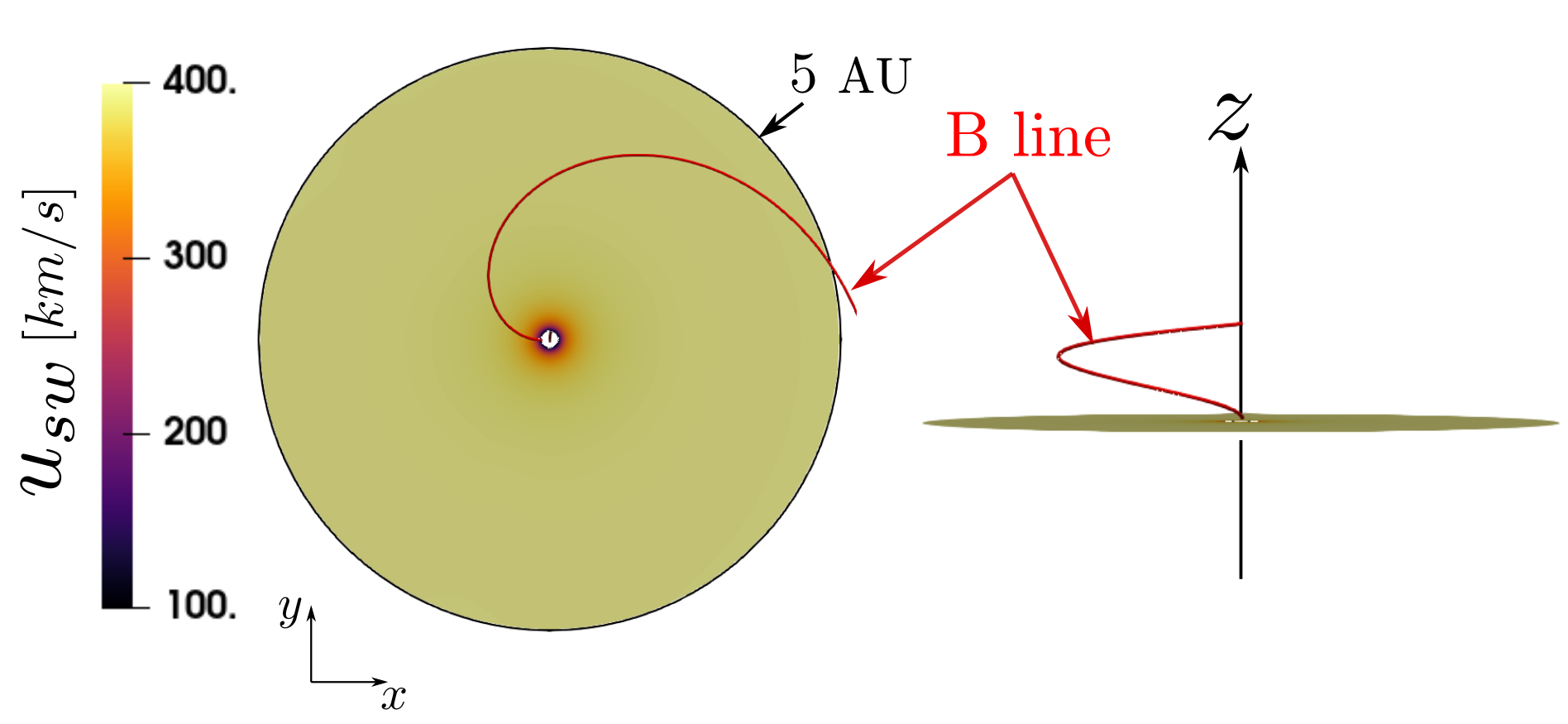}
    \caption{\label{fig:MHD_slices} Inner region of the MHD simulation. Top and side view of the heliographic equatorial plane, colored with the velocity of the solar wind $u_{sw}$ after steady state has been reached. An anti-sunward magnetic field line emanating from the domain at $24^{\circ}$ latitude is also shown in red.}
\end{figure}
To prevent an excessive longitudinal and latitudinal stretching of the grid toward the outer boundary, we also adopted a constant grid-stretching factor of 1.02 in the radial direction. 
For the purpose of this paper, we assumed the internal magnetic field of the Sun to be a centered dipole of strength $D=10^{-4}{\rm T}\: R_{\odot}^3$ ($R_{\odot}$ is the radius of the Sun), whose axis is aligned with the rotation axis (the $z$-axis).
 The plasma between the Sun and the inner boundary of the simulation domain was assumed to rotate rigidly. Consequently, at the inner boundary, the components of the fluid velocity tangential to the surface are given by $\Omega\:\hat{z}\times {\bf r}$, where $\hat{z}$ is the unit vector along the $z$-axis, and $\Omega=2\pi/(30 {\: \rm d})$ is the angular rotation velocity of the Sun.  
 The radial component of the fluid velocity obeys the free stream condition $\partial u_r/\partial r =0$, while the plasma temperature and the mass density were kept at $2 {\rm \ MK}$ and $1.2\times 10^{-20}{\rm \ kg/m^3}$, respectively. At the outer boundary, we adopted  $\partial/\partial r =0 $ for all quantities. 
As in the simulations of planetary magnetospheres of \cite{Griton_etal_2018}, and more generally, in simulations in which the magnetic field varies by several orders throughout the simulation domain, we conveniently split the magnetic field $\mathbf{B}$ into a prescribed potential background field $\mathbf{B}_0$ (in this case, the unperturbed solar dipole field) and a residual field $\mathbf{B}_1$  (the advantages of the decomposition were discussed by \cite{Tanaka_1994}). 

With this setup, the wind emanating from the inner boundary rapidly reaches a steady and supersonic radial speed of about 400 km/s, and the magnetic field lines spiral away from the inner boundary, as shown in Fig. \ref{fig:MHD_slices}. The equatorial current sheet and a nonconstant wind speed cause the resulting field structure to be similar but not identical to the analytic Parker spiral field  model. The density and the magnetic field in the simulated wind are lower than average values, but are still representative of the real wind. The wind parameters are given in Table  \ref{tab:swprop}, including the variation scale of the magnetic field strength $L_{\rm B}\equiv -(d \ln B/ds)^{-1}$, where $s$ is the spatial coordinate along the magnetic field line. 
\begin{table}[ht]
\centering
   \caption{\label{tab:swprop} Wind parameters at $r=1\:{\rm AU}$ on the field line at $24^\circ$ latitude of Fig. \ref{fig:MHD_slices}. Here, $\mu_0$ is the permeability of vacuum.}
    \begin{tabular}{ll}
        \hline \hline \\
        Magnetic field strength $B$                           & $ 0.89\: {\rm nT}$  \\
        Wind speed $u_{\rm sw}$                                        & $418\:{\rm km/s}$ \\
        Sound speed $c_{\rm s}=(\tfrac{5}{3}p/\varrho)^{1/2}$ & $48.2\:{\rm km/s}$ \\%
        Alfvén speed $c_{\rm A}=(B^2/\mu_0\varrho)^{1/2}$     & $71.9\:{\rm km/s}$\\%
        Plasma beta $\beta=p\:2\mu_0 /B^2$                                   & $0.54$\\%
        Magnetic field scale length $|L_{\rm B}|$     & $0.93\: {\rm AU}$ \\
       \hline \hline
    \end{tabular}
\end{table}
\subsection{\label{subsec:GCA}Particle dynamics: Guiding center approximation}
In the GCA, 
the relativistic motion of a test particle of rest mass $m$ and electric charge $q$ is described by the following set of equations (see, e.g., \cite{Ripperda_etal_2017a}):
\begin{eqnarray}
\label{eq:gca1}
 \frac{\mathrm{d}\mathbf{R}}{\mathrm{d}t} &=& v_{\|}\mathbf{b} + \mathbf{v}_{\rm E} + \frac{\gamma m}{qB}\mathbf{b} \nonumber \\ 
 &\times& 
 \left[ 
 \frac{\mu_{\rm B}}{\gamma^2 m} \nabla B + \frac{v_{\|}}{\gamma}E_{\|}\mathbf{v}_{\rm E} +  v_{\|}^2\left(\mathbf{b}\cdot\nabla\right)\mathbf{b} \right.
 \nonumber \\
 &+& 
 \left. \vphantom{\frac{1}{2}} v_{\|}( \left(\mathbf{v}_{\rm E}\cdot\nabla\right)\mathbf{b} 
+ \left(\mathbf{b}\cdot\nabla\right)\mathbf{v}_{\rm E}) + \left(\mathbf{v}_{\rm E}\cdot\nabla\right)\mathbf{v}_{\rm E} 
 \right] \\
\label{eq:gca2}
\frac{\mathrm{d} (\gamma v_{\|})}{\mathrm{d}t}  &=&  
 \frac{q}{m}E_{\|} - 
 \frac{\mu_{\rm B}}{\gamma m}\mathbf{b}\cdot\nabla B \nonumber \\
    &+& \gamma\mathbf{v}_{\rm E}\cdot
    \left[
      v_{\|}\left(\mathbf{b}\cdot\nabla\right)\mathbf{b}+
      \left(\mathbf{v}_{\rm E}\cdot\nabla\right)\mathbf{b}
    \right] \\ 
\label{eq:gca3}\frac{\mathrm{d}\mu_{\rm B}}{\mathrm{d}t}  &=& 0.  
\end{eqnarray}
In the above equations, the subscripts $\parallel$ and $\perp$ indicate projections parallel and perpendicular with respect to $\mathbf{B}$. $\mathbf{R}$ is the position of the particle guiding center, and ${\bf v}$ the guiding center velocity, $\gamma \equiv 1/\sqrt{1-v^2/c^2}$ is its Lorentz factor (with $c$ the speed of light). Hence, hereafter, $\mathcal{E}=(\gamma-1)mc^2$ denotes the relativistic kinetic energy of the particle. $\mu_{\rm B}\equiv \tfrac{1}{2}m\gamma^2v_{\perp}^2/B $ is its magnetic moment, and $\mathbf{v}_{\rm E} \equiv \mathbf{E}\times \mathbf{b}/B$ is the E-cross-B drift velocity, where $\mathbf{b}\equiv{\bf B}/B$ is the unit vector along $\bf B$.  In ideal MHD, the electric field is given by $\mathbf{E} = -\mathbf{u}\times \mathbf{B}$ (ideal Ohm's law), where ${\bf u}$ is the velocity of the fluid. However, since the electric field enters the system of MHD equations through $\partial {\bf B}/\partial t=-\nabla\times{\bf E}$, an  irrotational component can be added to the electric field without affecting the MHD flow. 
The contribution of the charge neutralizing electrostatic field due to the electron fluid $-\nabla p_e /(en_e)$ can be considered and the nonideal two-fluid Ohm law $\mathbf{E} = -\mathbf{u}\times \mathbf{B} - \nabla p_{\rm e}/(en_{\rm e})$ can be used, where $e$ is the positive elementary charge, $p_{\rm e}$ is the electron pressure, and $n_{\rm e}$ is the electron number density. 
We assumed a fully ionized electron-proton plasma so that we can write $\nabla p_{\rm e}/n_{\rm e}=2\kappa(1+\kappa)^{-1}\nabla p/n$, where $\kappa\equiv T_{\rm e}/T_{\rm p}$ is the electron-to-proton temperature ratio, which we set to $\kappa=1$. 
The expression for the electric field used in the particles simulations is therefore 
\begin{equation*}
{\bf E} = -\mathbf{u}\times \mathbf{B} - \frac{1}{en}\nabla p.
\end{equation*}
The corrected field is not irrotational in general. However, since $p=p_0$ and $n=n_0$ over the whole inner boundary, since all fluid particles in the domain emanate from the inner boundary, and since $p/n^{\Gamma}$ is constant along streamlines ($\Gamma$ is the adiabatic index), it follows that $p=p(n)$ and therefore $\nabla \times (\nabla p / n )= 0$, as required.
The wind simulation was compatible with the two-fluid Ohm law, and we therefore used it to evaluate the electric field in the simulated wind. We note, however, that 
the large-scale electrostatic field is generally of minor importance for SEP particles as the difference of electrostatic potential between the corona and $r=1\:{\rm AU}$ is only about a few hundred Volt, which becomes relevant for SEP particles below a few keV.
We note that the above GCA equations are for nonrelativistic flows with $|\mathbf{u}|\ll c$, a condition that is largely satisfied in the solar wind (relativistic flows were considered in \cite{Ripperda_etal_2017a}).  

The positions and velocity of test particles were advanced in time according to the GCA equations (\ref{eq:gca1})-(\ref{eq:gca3}) and the fields from the MHD simulation. We used a trilinear interpolation from the MHD grid to evaluate the fields at the particle position. Particles were advanced in time using the third-order prediction-correction time-integration scheme by \cite{Mignone_etal_2023}, which we briefly outline below. 

We set $\mathbf{X}\equiv\{\mathbf{R},\gamma v_{\|}\}$ and denote $\mathbf{F}$ to be the right-hand-side terms of the equations (\ref{eq:gca1})-(\ref{eq:gca2}). We then write the GCA equations as  
\begin{equation}
    \label{eq:all_gca_in_one}
    \frac{\mathrm{d}\mathbf{X}}{\mathrm{dt}}=\mathbf{F}\left({\bf X}\right).
\end{equation}
Integration of (\ref{eq:all_gca_in_one}) from a time level $t^n$ to $t^{n+1}=t^n+\Delta t^n$ was made by a predictor step followed by a corrector step, 
\begin{eqnarray}
\label{eq_xstar}
\mathbf{X}^* &=& \mathbf{X}^n \nonumber \\
 &+& \Delta t^n \left[ \left( 1 + \frac{\tau^n}{2} \right) \mathbf{F}^n-  \frac{\tau^n}{2} \mathbf{F}^{n-1} \right] 
+ \mathcal{O}(\Delta t^3) \\
\label{eq_xn1}
\mathbf{X}^{n+1} &=& \mathbf{X}^n \nonumber \\
&+& \frac{\Delta t^n}{6(1+\tau^n)}  
\left[ 3 \left(\mathbf{F}^n + \mathbf{F}^{*}\right)
+ 4 \tau^n \left(\mathbf{F}^{n}+\frac{\mathbf{F}^{*}}{2}\right) 
\right.
\nonumber\\ 
&+& (\tau^n)^2 (\mathbf{F}^{n} - \mathbf{F}^{n-1}) 
\left. \vphantom{\frac{1}{2}}\right] + \mathcal{O}(\Delta t^4),
\end{eqnarray}
where $\mathbf{X}^n = \mathbf{X}(t^n)$, $\mathbf{F}^n = \mathbf{F}({\bf X}^n)$ and $\tau^n\equiv\Delta t^n/\Delta t^{n-1}$. We highlight that the factor 3 in (\ref{eq_xn1}) is missing in the corresponding equation (15) of \cite{Mignone_etal_2023}.  
At $t=0$, or after a collision, the unknown quantities at $t^{n-1}$ were computed with a simple backward Euler step: $\mathbf{X}^{n-1} = \mathbf{X}^n - \Delta t^n\mathbf{F}^{n}$.

\subsection{Pitch-angle scattering \label{sec:pitch}}
In the solar wind, the mean free path for Coulomb collisions is about 1 AU for thermal ions and electrons. Since it grows as $v^4$, Coulomb collisions are completely ineffective for energetic particles. However, energetic particles are diffused in velocity space by turbulent fluctuations of the electromagnetic field of the wind \citep{Parker_1965}. 
A quasi-linear theory of the pitch-angle diffusion of charged particles in a weak turbulent magnetic field superposed on a strong guiding field, based on the Fokker-Planck equation, was first developed by \cite{Jokipii_1966, Jokipii_1968}. For power-law turbulence and nearly isotropic distributions, the quasi-linear phase-space diffusion coefficient takes the general form 
\begin{equation}
D_{\upmu \upmu}(\upmu) = \frac{1}{2} \left< (\Delta \upmu)^2 / \Delta t    \right> = h(\upmu) \frac{1-\upmu^{2}}{2} \label{eq_D_mumu},
\end{equation}
where $\upmu \equiv  \cos \alpha $, and $\alpha$ is the angle between the particle velocity ${\bf v}$ and the magnetic field $\bf B$, and $h(\upmu)$ is a model-dependent function (see \cite{Schlickeiser_1989,Wijsen_etal_2019}). A parallel mean free path $\lambda_{\|}$ can be computed from the diffusion coefficient (\cite{Jokipii_1966,Hasselmann_1968,Bieber_1994}) via
\begin{equation}\label{eq:lamb_par}
    \lambda_\parallel = \frac{3v}{8}\int_{-1}^{+1} \frac{\left(1-\upmu^2 \right)^2}{D_{\upmu\upmu}(\upmu)}\mathrm{d}\upmu.
\end{equation}
For constant $h(\upmu)=h_0$ in (\ref{eq_D_mumu}), $h_0=v/\lambda_\parallel$, which can thus be interpreted as the collision frequency for $\upmu=1$ (or $\alpha=0^\circ$) particles.   
In equation (\ref{eq:lamb_par}) the parallel mean free path depends on the velocity. However, since $D_{\upmu\upmu}(\upmu)$ is proportional to the collision frequency, $v/\langle D_{\upmu\upmu}\rangle$ can be interpreted as a measure of the distance between scattering centers, and $\lambda_{\|}$ is expected to be only weakly dependent on the particle energy. At 1 AU, the observed mean free paths of protons and electrons are in the range 0.1 to 1 AU for a large interval of rigidities from $10^{-4}$ MV to $10^4$ MV  \citep{Palmer_1982,Bieber_1994}.

Diffusion is generally thought to occur through the accumulation of small deviations. This can be simulated numerically by applying a Fokker-Planck diffusion operator to an initial distribution of particles. 
We adopted a different approach in which test particles underwent hard-sphere-type collisions as in \cite{Marsh_etal_2013} and \cite{Dalla_etal_2013,Dalla_etal_2015}.  
In practice, for a particle that covered a distance $\Delta s = \int_0^{\Delta t} v_\parallel dt$ during the time step $\Delta t$, the collision probability is given by
\begin{equation}
    p(\Delta s) = 1-\exp(-\Delta s/\lambda_{\parallel}). 
\end{equation}
At each time step, a random number $a$ is generated in the interval $[0,1]$. If $a<p(\Delta s)$, the particle undergoes an elastic collision. In the case of a collision, the post-collision pitch angle $\alpha=\arccos(\upmu)$ (in the frame of the scattering centers) is selected to give a uniform distribution in $\upmu$  
\begin{equation}
    \upmu = 1-2a,\; \text{$a$: random number}\in[0,1]. \label{eq_alpha_iso}
\end{equation}
Given the scattering law (\ref{eq_alpha_iso}), all directions in 3D space are equally probable. The law does not ensure isotropy, however. The reason is that since the particle guiding centers move along the magnetic field at an angle-dependent speed $\upmu v$, the scattering centers do not receive an isotropic distribution  while emitting an isotropic distribution under the action of the scattering law (\ref{eq_alpha_iso}). As a result, collisions tend to overpopulate the directions perpendicular to the magnetic field. To force isotropy in the highly collisional limit, the post-collision angular distribution  of the particles hitting the scattering centers must be equal to the pre-collision distribution. 
The latter is obviously proportional to the rate at which particles from an isotropic distribution hit the scattering centers during their one-dimensional displacement along the magnetic field. The number $\mathfrak{n}$ of scattering events per time unit observed at a scattering center with a pitch angle $\alpha$ such that $\cos\alpha > |\upmu|$ is given by \citep[see Sect. II.C in][]{Pantellini_2000}
\begin{equation}
 \mathfrak{n}(\upmu) =\mathfrak{n}_0 -2\int_0^{\upmu} \mathfrak{n}_0 \tilde\upmu\: d\tilde\upmu = \mathfrak{n}_0 (1-\upmu^2)\label{eq:nu_aniso},
\end{equation}
where $\mathfrak{n}_0$ is the cumulated number of events for $\upmu$ in the range $[-1,1]$. 
In this case, $\mathfrak{n}$ varies linearly with respect to $\upmu^2$ instead of $\upmu$ ,
and the scattering law (\ref{eq_alpha_iso}) has to be modified accordingly,  
\begin{equation}
    \upmu = \pm\sqrt{a},\; \text{$a$: random number}\in[0,1] \label{eq:alpha_aniso},
\end{equation}
with the plus and minus sign being equally probable. The mean free path $\lambda_\parallel$ is the distance between the scattering centers and is therefore identical for all particles, that is,  $\lambda_\parallel=v_\parallel/\upnu(\upmu)=v/\upnu_0$, where $\upnu(\upmu)=\upnu_0\upmu$ is the particle collision frequency. From (\ref{eq:lamb_par}) and the general form of $D_{\upmu\upmu}$ (\ref{eq_D_mumu}), it follows that the corresponding Fokker-Planck diffusion coefficient is $D_{\upmu\upmu}=\upnu_0(1-\upmu^2)/2$, which is the form used by \cite{Zaslavsky_etal_2024}. Hence, differences between our results and those of \cite{Zaslavsky_etal_2024} are expected for a scattering model other than (\ref{eq:alpha_aniso}). 
Since there are no stringent reasons for the distribution of test particles to be isotropized by the scattering, we adopt the scattering law (\ref{eq_alpha_iso}) for which all post-scattering directions are equally probable in the following, but we retain that the diffusion in phase-space is not of the Fokker-Planck type.

\section{\label{sec:results} Results and discussion}
\begin{figure}[ht]
\centering
 \includegraphics[width=0.5\textwidth]{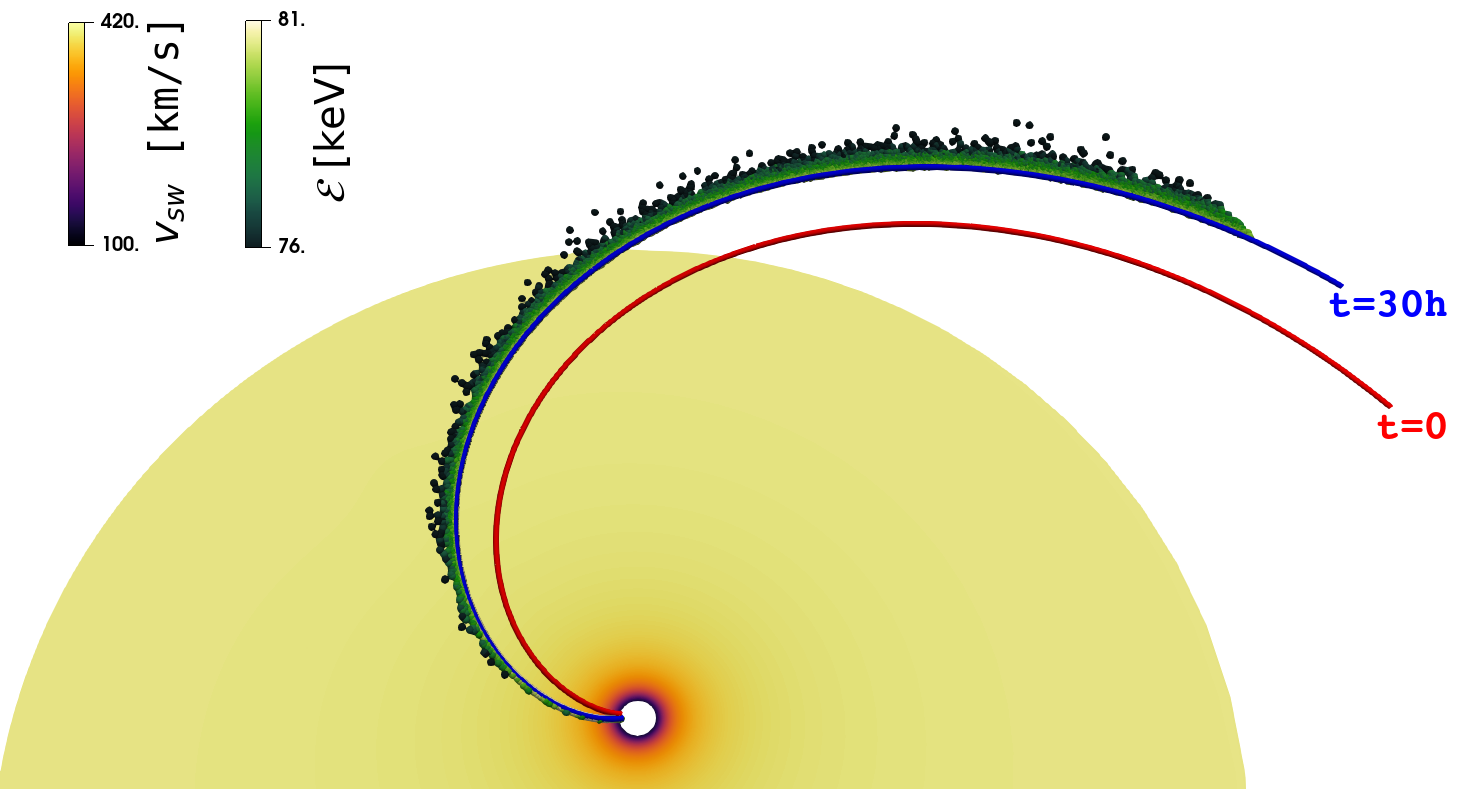}
 \caption{\label{fig:particles_inSW} Top view of the spatial distribution of the electrons at $t=30\:{\rm h}$.
 In blue we plot the field line on which the particles were injected. In red we plot the same field line $t=30\:{\rm h}$ earlier, at the time of the injection of the oldest particles. 
 }
\end{figure}

\begin{figure}[h]
\centering
 \includegraphics[width=0.45\textwidth]{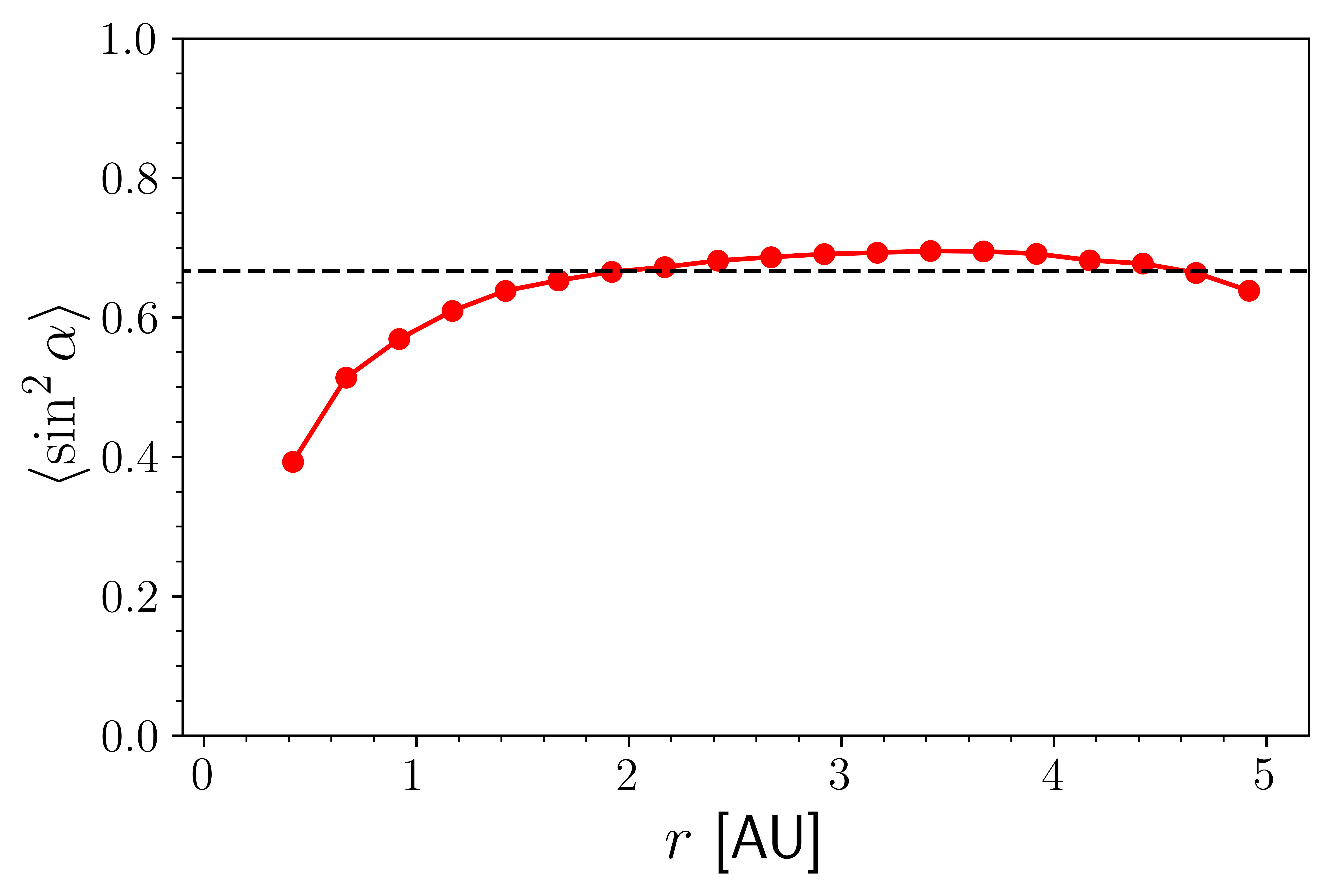}
 \caption{\label{fig:sin2alpha} Simulation with $\lambda_{\|} = 0.5\:{\rm AU}$. Pitch-angle distribution $\langle \sin^2 \alpha \rangle $ as a function of heliocentric distance. The dashed horizontal line corresponds to the isotropic distribution for which $\langle \sin^2 \alpha \rangle=2/3 $.}
\end{figure}

\begin{figure}[h]
\centering
 \includegraphics[width=0.45\textwidth]{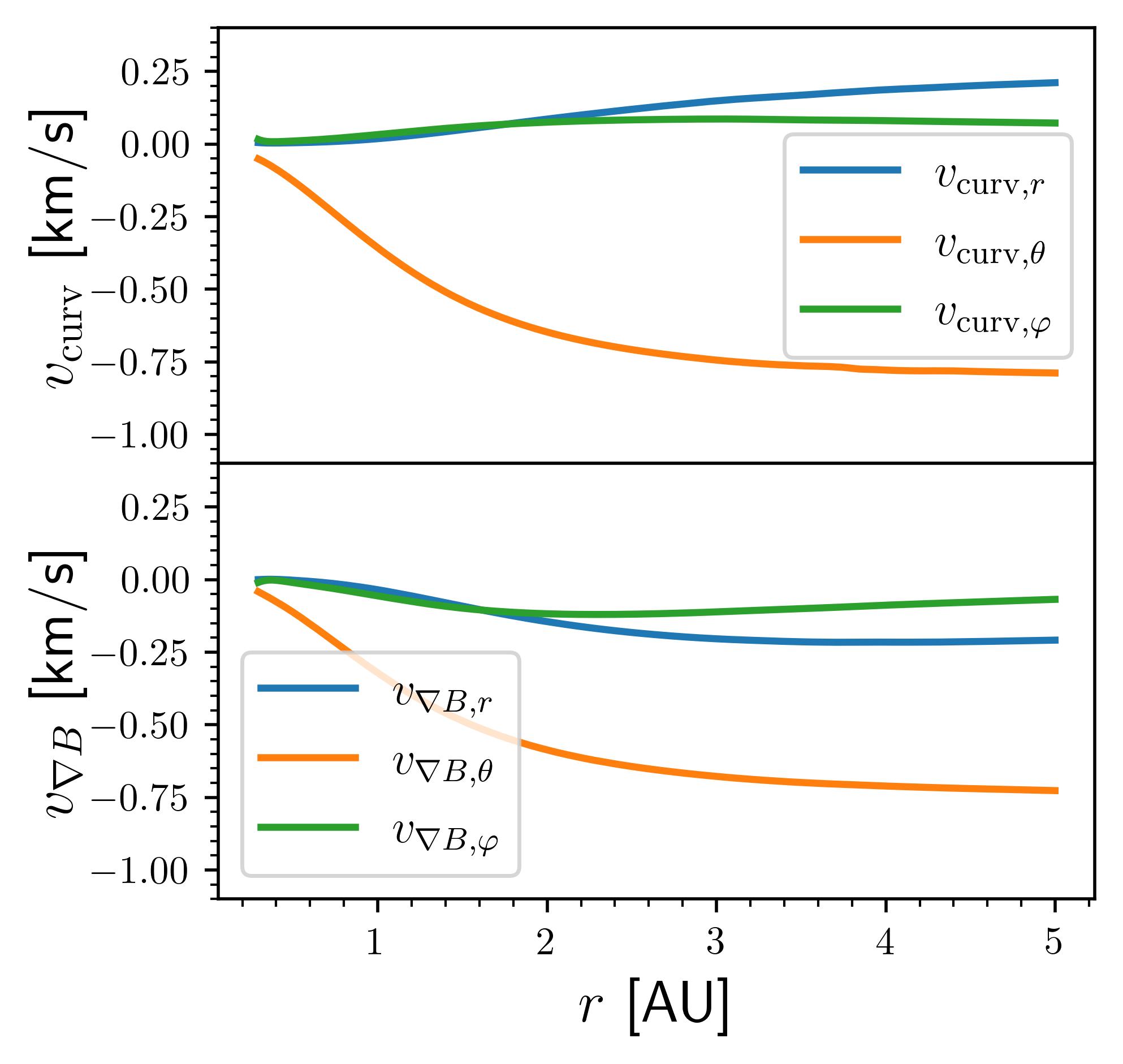}
 \caption{\label{fig:vdrift} Simulation with $\lambda_{\|} = 0.5$ AU.  Radial variations in the components of the curvature drift (top panel) and magnetic gradient drift (bottom panel) for a $81 {\rm keV}$ electron with a constant pitch $\sin^2 \alpha = 0.64 $.}
\end{figure}

Initially, $10^3$ mono-energetic electrons were impulsively injected with $81\:{\rm keV}$ energy and $\alpha = 0$ at $r=0.28\:{\rm AU}$ (corresponding to 60 $R_\odot$) and $24^{\circ}$ latitude on the corotating magnetic field line shown in Fig. \ref{fig:MHD_slices}. The choice of $\alpha=0$ was arbitrary and did not significantly impact the results because the mirror force naturally drives the perpendicular velocity toward zero as particles propagate outward in the decreasing magnetic field. We set $\lambda_{\parallel}$ to three different values \{0.1, 0.5, and 1\} AU in three different simulations, in conformity with the consensus on the mean free path of high-energy particles near 1 AU \citep{Palmer_1982,Bieber_1994}. In the following, except in Sect. \ref{subsec:pda}, we focus exclusively on the case of $\lambda_{\parallel}= 0.5\:{\rm AU}$.
We adopted a constant time step $\Delta t=0.54\:{\rm s}$ to integrate the GCA equations and  perform scattering in the inertial frame so that the particles did not experience adiabatic cooling (cf. Sect. \ref{subsec:ac}).
Particles that left the spherical shell $[0.28,5]\:{\rm AU}$\footnote{The outer boundary was set to 2 AU and 10 AU for $\lambda_{\|}=0.1\:{\rm AU}$ and $\lambda_{\|}=1\:{\rm AU}$, respectively.} were reinitialized (reborn) on the same magnetic line with the same initial velocity during $150\:{\rm h}$, which was long enough for each particle to be reborn $20$ times on average. To check for a steady state, we monitored the time evolution of the mean heliocentric distance of the particles. When this quantity stabilized, we considered the system to have reached steady state. This occurred around $t=8.3\:{\rm h}$. We were primarily interested in the steady-state distribution of particles, so that in the following, we only consider particles that were injected after $t=8.3\:{\rm h}$. To make statistics, we accumulated snapshots of the system every 100 time steps ($=54\:{\rm s}$, roughly one-tenth of the collision time), which is equivalent to an ensemble of some $9.4\,10^6$ electrons.   
As an example, a snapshot of the spatial distribution and the energy of the electrons at $t\sim 30 \:{\rm h}$ is reported in Fig. \ref{fig:particles_inSW}. 
The figure shows that particles are dominated by a ${\bf v}_{\rm E}$ drift in the azimuthal direction (toward positive $\varphi$). Even if this drift, which scales with the solar wind velocity ($\sim 400$ km/s), is much faster than both the curvature and gradient drifts (see below), it does not contribute  to the particle energy loss from 81 to 76 keV. The reason is that the ${\bf v}_{\rm E}$ drift is  by definition transverse to the electric field and only displaces particles on equipotential lines.
Particles may change their energy when they move along the magnetic field line (first term on the right-hand side of Eq. (\ref{eq:gca1}) because of the nonzero field-aligned electrostatic field $E_\parallel$, but as already mentioned, this field is weak, especially at heliocentric distances beyond the sonic point. We considered the two dominant drifts (for particles with $v\gg u_{\rm sw}$) from the square brackets on the right-hand side of (\ref{eq:gca1}). These are the curvature drift velocity 
\begin{equation*}
    {\bf v}_{\rm curv} = [\gamma m v_{\|}^2 /(q B)]\:{\bf b} \times ({\bf b}\cdot\nabla) {\bf b},
\end{equation*}
and the gradient drift velocity 
\begin{equation*}
    {\bf v}_{\nabla B} = [\gamma m v_{\perp}^2 /(2 q B)]\:{\bf b} \times \nabla {\rm ln} B.
\end{equation*} 
We anticipate that for a $\mathcal{E}=81\:{\rm keV}$ electron (or proton) in a $B=1\:{\rm nT}$ field and typical magnetic gradient scales of about $L_{\rm B}=1\:{\rm AU}$, the expected drift velocities are low as $v_{\rm curv} \sim v_{\nabla B} \sim \mathcal{E}/(eB L_{\rm B}) \approx 0.5\:{\rm km/s}$ . For a fixed kinetic energy (i.e., for a fixed $v$), both terms can be written in terms of $\sin^2 \alpha$ as  $v_\perp^2=v^2 \sin^2\alpha$ and $v_\parallel ^2=v^2(1-\sin^2 \alpha)$. 
Averages of $\sin^2\alpha $ as a function of heliocentric distance are reported in Fig. \ref{fig:sin2alpha} for the $\lambda_  \parallel=0.5\: {\rm AU}$ simulation. The figure shows that the pitch-angle distribution tends to evolve from beam-like to isotropic with increasing distance, indicating a decreasing efficiency of the magnetic focusing. The variations in the curvature and gradient drift with heliocentric distance clearly depend on the variation in $\alpha$ and both the variation in the magnetic curvature and the magnetic gradient in the plane perpendicular to ${\rm B}$. The radial profiles of all components of $ {\bf v}_{\rm curv}(r)$ and $ {\bf v}_{\nabla B}(r)$ 
are plotted in Fig. \ref{fig:vdrift} for a constant $\sin^2 \alpha = 0.64$ (average value from the curve in Fig. \ref{fig:sin2alpha}). The drifts are charge dependent (protons and electrons drift in opposite directions), and their main components along $\theta$ increase linearly for $r\lesssim 1.5\:{\rm AU}$ and become nearly constant for $r\gtrsim 2\:{\rm AU}$. In this configuration, both drifts correspond to a poleward motion of the electrons (toward decreasing $\theta$) and an equatorward motion of the protons. Both drifts are slow but perpendicular to the fast ${\bf v}_{\rm E}$
drift. 
In Fig. \ref{fig:particles_inSW}, black particles have experienced greater drift in longitude and latitude, and their energy loss is consequently greater. 

\subsection{Energy loss}
From an order-of-magnitude estimate of the terms in Eq. (\ref{eq:gca2}), it is clear that for 81 keV electrons, energy changes are essentially caused by the field curvature and field gradient terms. Assuming time steadiness, we write (\ref{eq:gca2}) in terms of the time evolution of the kinetic energy of a particle,      
\begin{equation}\label{eq:ek}
\frac{d\mathcal{E}}{dt}\simeq  m\gamma v_\parallel^2\; 
{\bf v}_{\rm E}\cdot({\bf b}\cdot\nabla){\bf b} \;+ \; m \gamma \frac{v_{\perp}^2}{2}\; 
{\bf v}_{\rm E}\cdot \nabla {\rm ln} B. 
\end{equation}
In this configuration, ${\bf v}_{\rm E}$ is oriented opposite to the curvature vector $\left({\bf b}\cdot \nabla \right){\bf b} $ and to the magnetic field gradient $\nabla {\rm ln} B$, so that (\ref{eq:ek}) describes a systematic loss of energy, regardless of whether the particle is moving away from or toward the Sun \citep[also see Fig. 2 in][]{Dalla_etal_2015}.

Replacing ${\bf v}_{\rm E}$ with ${\bf E \times b}/B$, we write  Eq. (\ref{eq:ek}) as 
\begin{equation}\label{eq:ek2}
\frac{d\mathcal{E}}{dt}\simeq  q {\bf v}_d \cdot {\bf E},
\end{equation}
where ${\bf v}_d = {\bf v}_{\rm curv} + {\bf v}_{\nabla B} $ is the total drift velocity. Equation (\ref{eq:ek2}) shows that energy loss is due to the electrons that drift along the electric field, whose main component is in the $-\theta$ direction. Protons also lose energy, as they drift in the opposite direction. It is evident that particles lose energy independently of their charge because the charge does not appear in Eq. (\ref{eq:ek}) \cite[see also][]{Dalla_etal_2015}.
\begin{figure}[h]
\includegraphics[width=.5\textwidth]{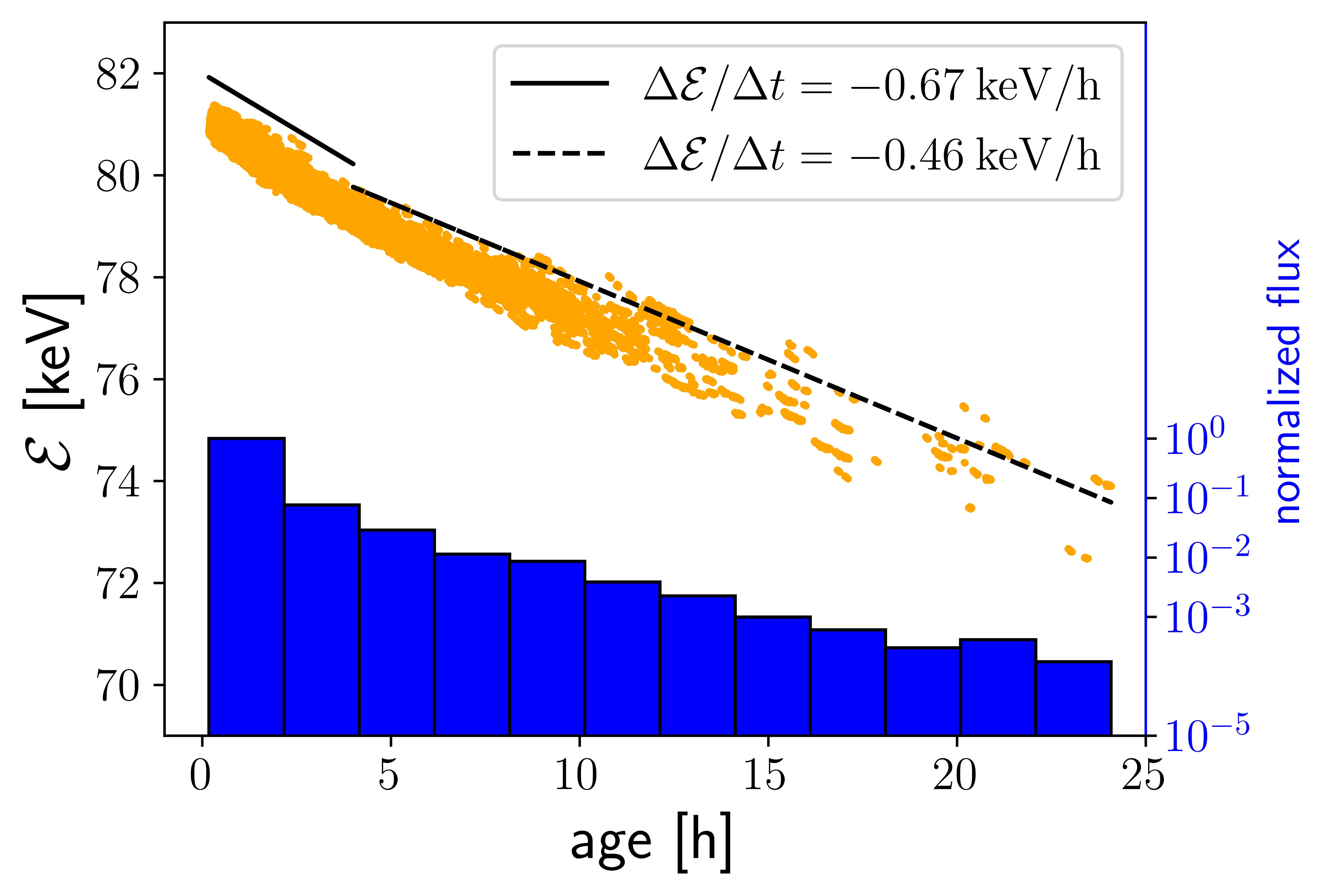}
\includegraphics[width=.45\textwidth]{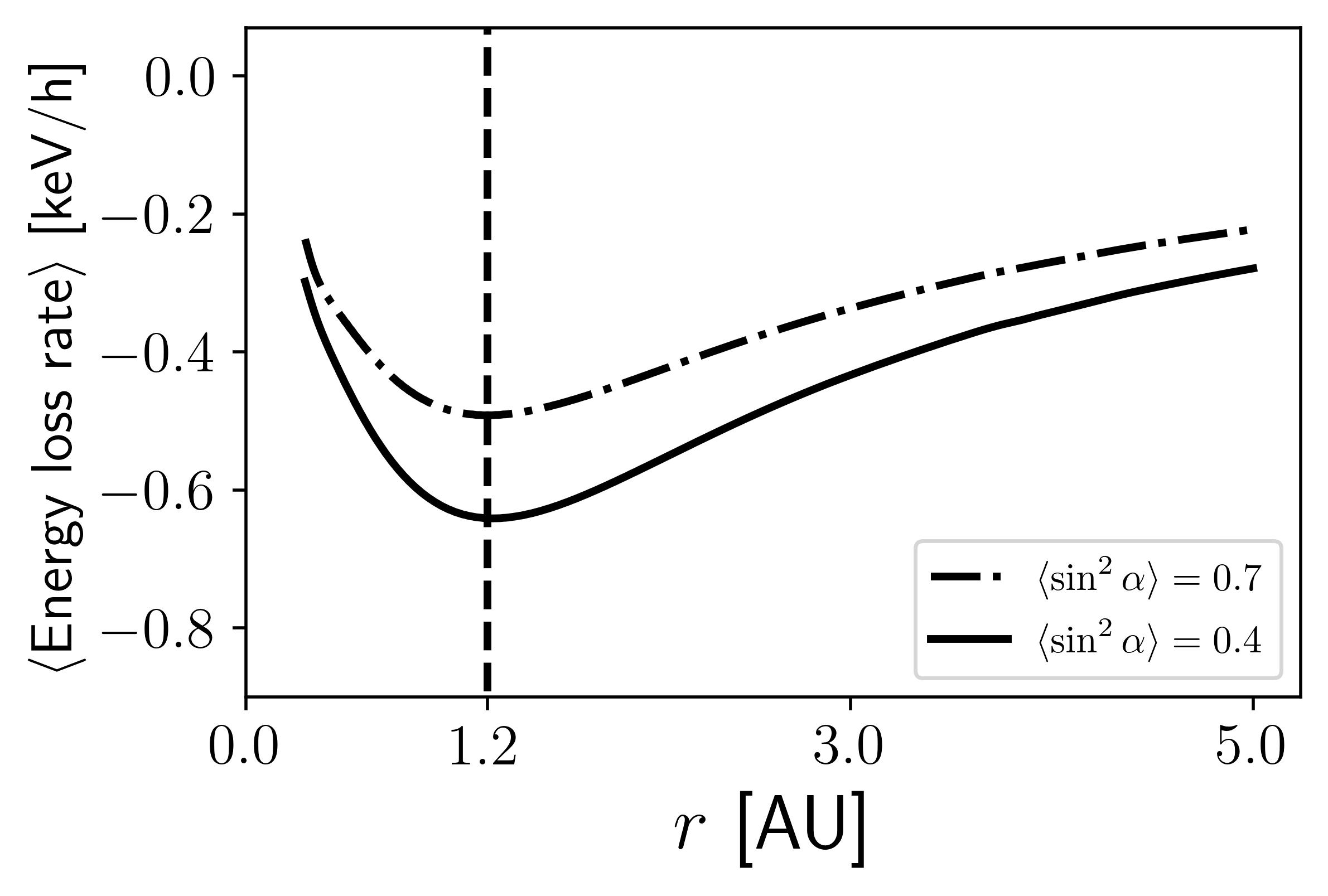}
\caption{\label{fig:ek_and_force_4}  Simulation with $\lambda_{\|} = 0.5\:{\rm AU}$. Top: Energy vs. age (time since injection) for electrons near $r=1$ AU. Linear fits of the decay rate for young (age < 4 h) and old (age$\ge$ 4 h) electrons are given. The normalized age distribution is also plotted. 
Bottom: Average energy-loss rate as a function of heliocentric distance $r$ given by Eq. (\ref{eq:ek}) for the extreme values of $\sin^2 \alpha$ observed in Fig. \ref{fig:sin2alpha}.} 
\end{figure}

The loss rate computed from the right-hand side of (\ref{eq:ek}) is reported in the bottom panel of Fig.\ref{fig:ek_and_force_4} for the two values of $\sin^2 \alpha $ that correspond to the extreme cases in  Fig. \ref{fig:sin2alpha}.
For the wind conditions in our simulation, the maximum loss rate occurs at $r=1.2\:{\rm AU}$. Toward increasing $r$, the loss rate drops because the curvature and magnetic gradient scale of the magnetic field lines increase. Toward decreasing $r$, the loss rate drops because of the decreasing ${\bf v}_{\rm E}$, as the fluid velocity and the magnetic field tend to align when they approach the inner boundary.

\begin{figure}[h]
\centering
\includegraphics[width=0.45\textwidth]{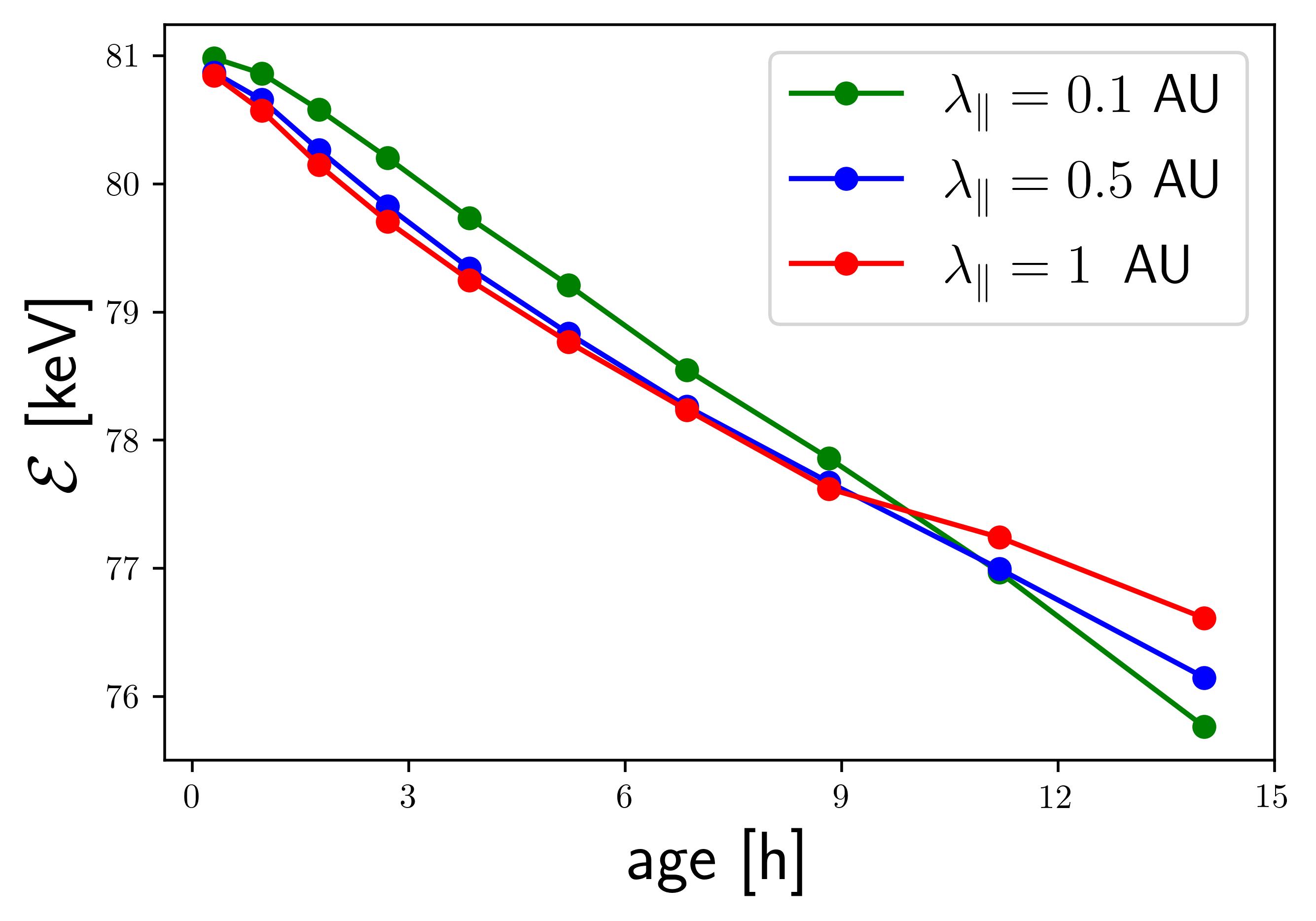}
\caption{\label{fig:ek_lam} Energy profiles as a function of age for electrons near $r=1$ AU and different values of $\lambda_{\|}$.}
\end{figure}

The loss of kinetic energy as a function of age (time since injection) for particles in the range $r\in[0.9,1.1]\:{\rm AU}$ is shown in the top panel of Fig. \ref{fig:ek_and_force_4}. The difference between young (age < 4 h) and old (age $\ge $ 4 h) can be understood in the line of the energy loss rate in the bottom panel of Fig. \ref{fig:ek_and_force_4}. Young particles did not have the opportunity to spend long periods of time at large heliocentric distances where the energy loss rate is substantially lower than near 1.2 AU, so that they lost energy faster than the old particles on average. 
As the energy loss rate is proportional to $v_{\|}^2$ and because collisions tend to isotropize the pitch-angle distribution, the energy loss rate might be thought to depend on the mean free path $\lambda_{\|}$. Figure \ref{fig:ek_lam} shows that the differences are relatively small for the $\lambda_{\|}$ values \{0.1, 0.5, and 1\}AU. 
As we showed in the bottom panel of Fig. \ref{fig:ek_and_force_4}, the energy loss rate depends on the pitch-angle distribution. The smaller $\langle \sin^2 \alpha \rangle$, the higher the energy loss rate, as for this wind configuration near 1 AU the curvature term is a more efficient cooler than the magnetic gradient term in (\ref{eq:ek}) . This explains the fast energy loss for young particles 
that spent most of their lifetime near 1 AU, where $\sin^2 \alpha $ is comparatively small. As particles grow older, the loss rate is reduced because old particles spent a significant fraction of their lives at large heliocentric distances, where $\sin^2 \alpha$ is largest.
The average displacement of a particle with respect to the observation point can be evaluated by assuming that the motion of the particle is a random walk. Its displacement with respect to the observation point at $r=1\:{\rm AU}$ might then be estimated to be about $\Delta s  \sim \sqrt{N}\lambda_\parallel$, where $N\sim vt/\lambda_\parallel$ is the number of collisions during the time $t$. For particles aged $t_{\rm a}=9{\rm h}$ ($t_{\rm a}$ is the time since injection) moving at $v=c/2$, the displacement $\Delta s \sim \sqrt{v t_{\rm a} \lambda_\parallel}$ is about $\{1.8; 4.0; 5.7\}\:{\rm AU}$ for $\lambda_\parallel=\{0.1;0.5;1\}\:{\rm AU}$, respectively. Hence, as loss rates decrease at more than $1\:{\rm AU}$ from the observation point (see the bottom panel of Fig. \ref{fig:ek_and_force_4}), loss rates are logically lower for old particles and a long collisional mean free path $\lambda_{\|}$. For comparison, the displacement is three times weaker and the age effect is negligible for particles that are only $1{\rm h}$ old.   

\subsection{\label{subsec:ac} Adiabatic cooling}

To distinguish the energy losses caused by the drifts from the energy losses caused by the scattering, we so far considered the case in which the scattering centers do not move with respect to the fixed inertial frame attached to the Sun. In this frame, the particle energy is conserved during collisions. However, when we assume that the scattering is due to the turbulence in the solar wind plasma, the appropriate frame for the scattering is the solar wind frame. Because of the combined action of scattering and magnetic focusing, the magnitude of the particle momentum $\Tilde{\mathbf{p}}= m \gamma \Tilde{\mathbf{v}}$ in this case changes in time according to \citep[see e.g.][]{ruffolo_1995}
\begin{equation}\label{eq:ruff}
\frac{1}{\Tilde{{p}}^2}\frac{{\rm d}\Tilde{{p}}^2}{{\rm dt}} = - \frac{u_{\rm sw}}{L_{\mathrm{B}}}\sin^2 \Tilde{\alpha},
\end{equation}
where the tilde quantities refer to the moving frame. 
In a radial field $B \propto 1/r^2$ and in the limit of frequent scattering (i.e., $\lambda_{\|} \ll |L_{\rm B}|$), the average particle position moves radially outward at a speed $u_{\rm sw}$, so that it sees a decreasing magnetic field and an average $ \langle L_{\rm B} \rangle  = r/2$. 
In the extreme case of frequent scattering, the pitch-angle distribution is close to isotropic, in which case $ \langle \sin^2 \Tilde{\alpha} \rangle = 2/3 $ and Eq. (\ref{eq:ruff}) reduces to 
\begin{equation}\label{eq:ruff1}
\frac{1}{\Tilde{{p}}^2}\frac{{\rm d}\Tilde{{p}}^2}{{\rm dt}} = - \frac{4}{3}\frac{u_{\rm sw}}{r}.
\end{equation}
Equation (\ref{eq:ruff1}) predicts a loss rate of $ -1.34 \% $ per hour, which in terms of kinetic energy loss rate corresponds to a loss rate of about -1 keV/h for a 81 keV electron. This is an upper estimate of the cooling rate, as for infrequent collisions, particles tend to accumulate near $\Tilde{\alpha}=0$, which reduces the loss rate to zero. 
\begin{figure}[t]
\centering
\includegraphics[width=0.45\textwidth]{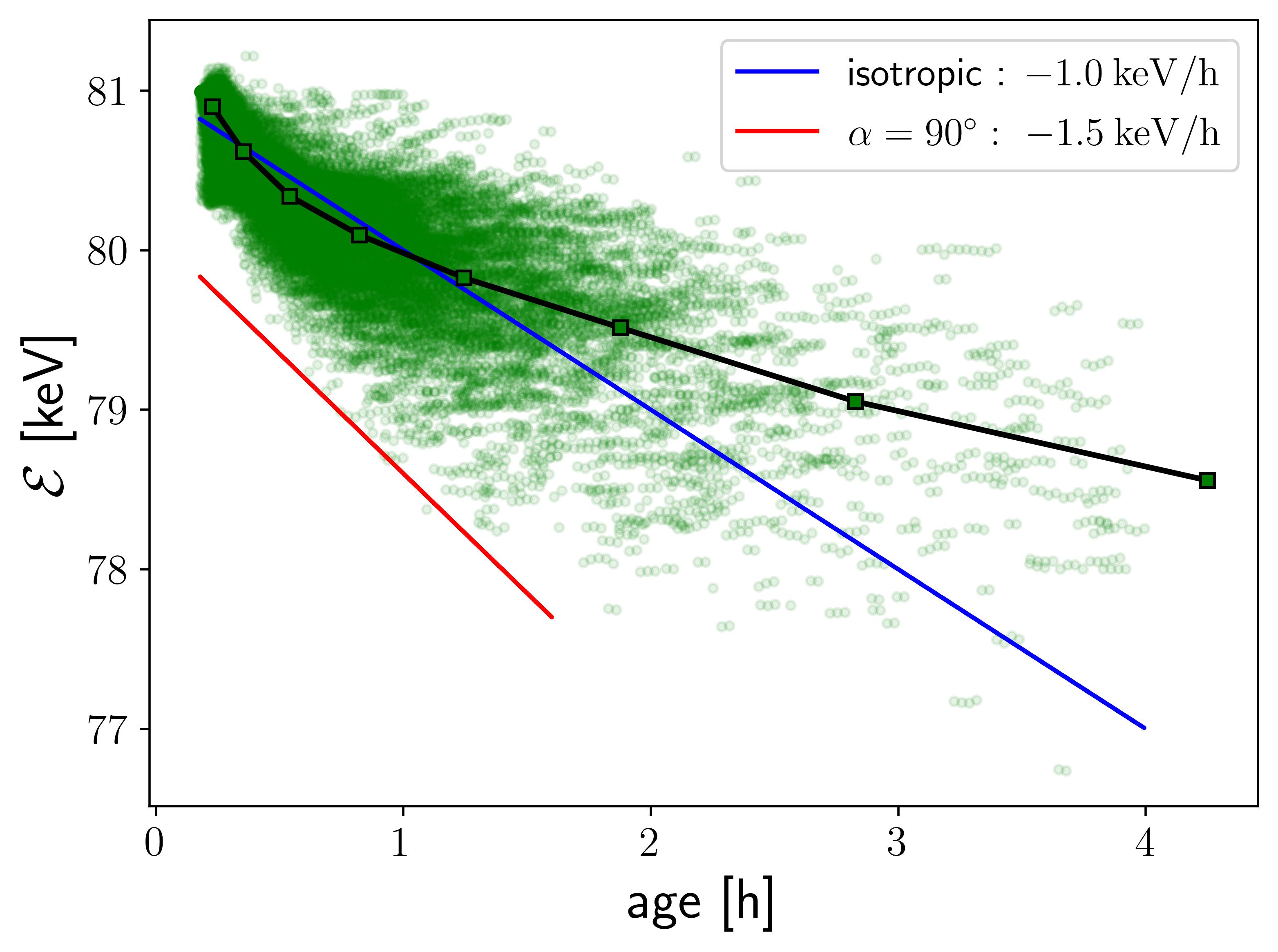}
\caption{\label{fig:adiabatic cooling} Energy vs. age near $r=1$ AU for electrons undergoing scattering in the solar wind frame with all drifts, except for $v_\parallel{\bf b}$, turned off. As in Fig. \ref{fig:ek_and_force_4}, the initial energy of the electrons is 81 keV, and the scattering mean free path is $\lambda_\parallel=0.5\:{\rm AU}$. The black line shows the average evolution. The slopes corresponding to the theoretical loss rates from the expression (\ref{eq:ruff1}) for a radial magnetic field and a wind velocity $u_{\rm sw}=418\:{\rm km/s}$ are also shown.}
\end{figure}
The energy loss that arises from the movement of the scattering center with the solar wind is shown in Fig \ref{fig:adiabatic cooling}. The only other difference with respect to the simulation of  Fig. \ref{fig:ek_and_force_4} is that solely the drift parallel to ${\bf B}$ is retained in the GCA equations. The large dispersion in energy clearly arises because all pitch angles $\alpha$ in the range 0 to $180^{\circ}$ are accessible, with each value of $\alpha$ giving a different cooling rate, as expressed by Eq. (\ref{eq:ruff}). Figure \ref{fig:adiabatic cooling} shows that the adiabatic loss rate continuously decreases with age, for instance, from -1.8 keV/h for particles aged 0.3 h to -0.6 keV/h for particles aged 1 h and -0.3 keV/h for particles aged 4 h. The reason is that the average heliocentric position $\langle r \rangle $ of very young particles is about the $r_0$ (observer's position) and $\sqrt{v t_{\rm a} \lambda_{\|}}$ for old particles aged $\gg r_0^2 / (v_{\rm sw} \lambda_{\|})$. Setting $r=\langle r \rangle $ in Eq. (\ref{eq:ruff1}), we obtain a loss rate $\sim v_{\rm sw} / r_0 $ for young particles and an age-dependent loss rate $\sim v_{\rm sw} / \sqrt{v t_{\rm a} \lambda_{\|}} $ for old particles. 

\begin{figure}[t]
\centering
\includegraphics[width=0.45\textwidth]{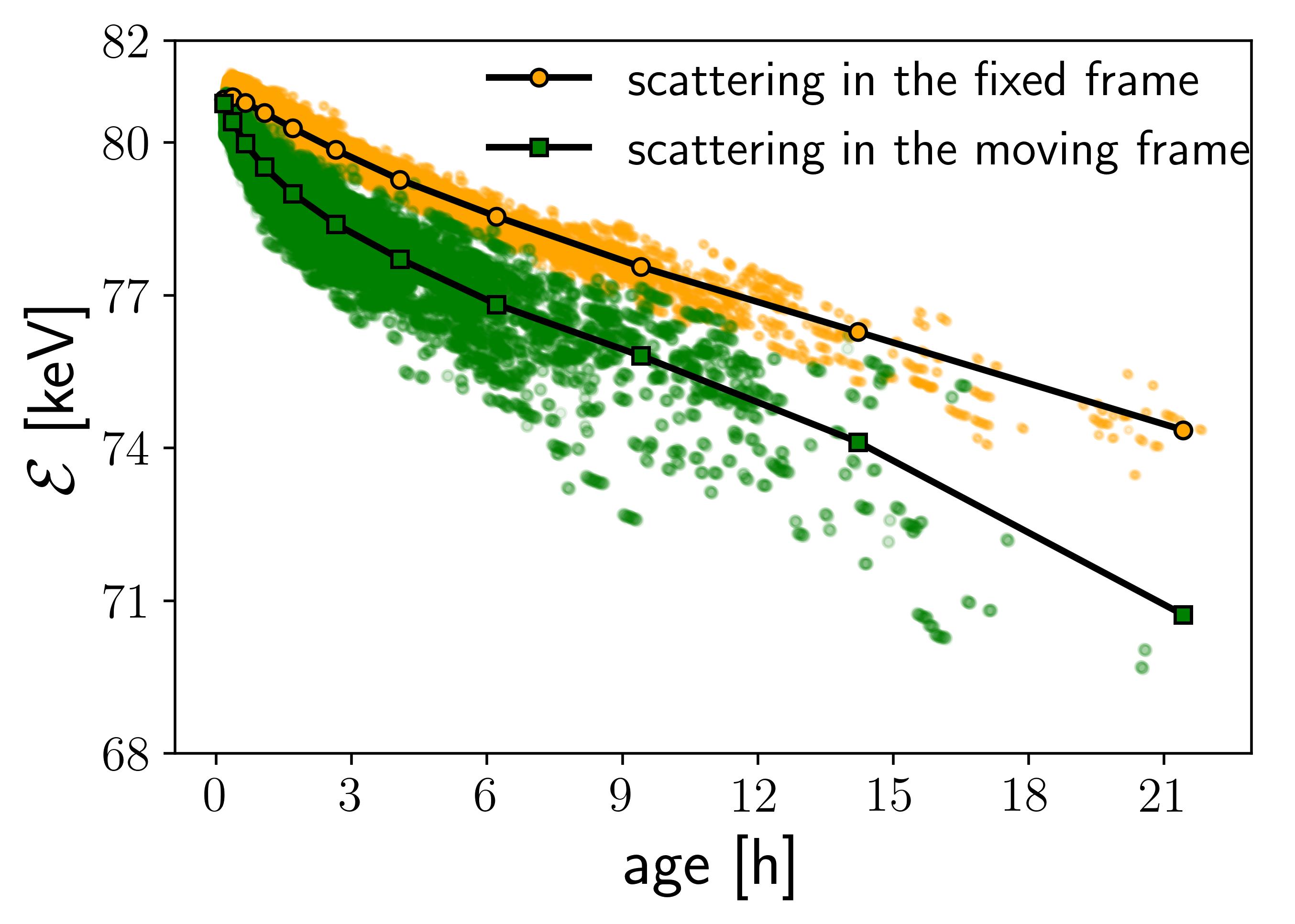}
\caption{\label{fig:ac+drift} Simulations with $\lambda_{\|} = 0.5 $ AU. The energy as a function of age near $r=1$ AU is shown considering only drift effects (orange dots) and considering both drifts and adiabatic cooling effects (green dots). The average profiles are also plotted for both cases.}
\end{figure}

Figure \ref{fig:ac+drift} illustrates the contribution of  the adiabatic cooling to the global energy loss rate for the case $\lambda_{\|}=0.5$ AU. The orange dots are from Fig. \ref{fig:ek_and_force_4} without adiabatic cooling particles (scattering in the fixed frame). The green dots correspond to the same simulation, except that the scattering centers are in the wind frame instead of the fixed frame.  As already suggested by Fig. \ref{fig:adiabatic cooling}, the adiabatic cooling is the dominant cooling mechanism for young particles. For particles aged 0.3 h, it  contributes -1.8 keV/h to the total loss rate of -2.1 keV/h. For particles older than -1.5 h, the contribution of the adiabatic cooling is small, which is confirmed by the fact that the energy versus age profile for the green dots is only marginally steeper than for the orange dots. As an example, at age 9 h, the cooling rates are about -0.4 keV/h and -0.3 keV/h, respectively.  
 
\subsection{\label{subsec:pda}Pitch-angle distribution}
In the case of a transient injection, for example, due to a solar particle event, the flux of 81 keV electrons at 1 AU reaches its peak $\sim 10\:{\rm min}$ after injection. As the peak intensity rapidly decreases on a timescale of about 10 to 20 minutes, only electrons younger than 20 min are representative of electrons from a solar particle event at the time of strongest particle flux at 1 AU. Retaining particles aged 20 minutes or less is equivalent to measuring the flux of particles from a solar event at 1 AU over a 10-minute interval.

\begin{figure}[h]
\includegraphics[width=0.45\textwidth]{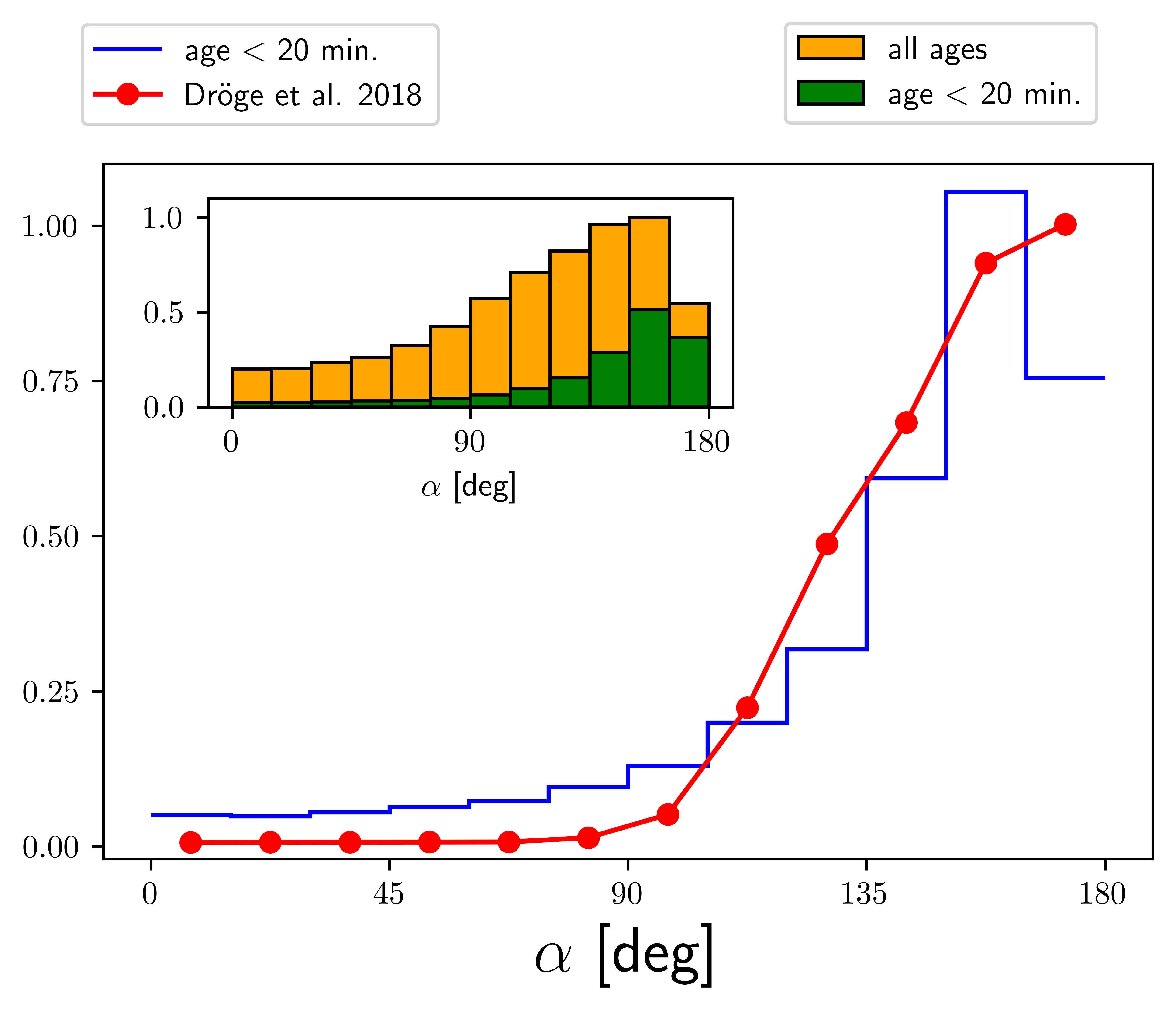}
 \caption{\label{fig:pitch_angle_dist_D} Normalized pitch-angle distribution for 81 keV electrons younger than 20 min with $\lambda
 _{\|} = 0.5 \:{\rm AU} $ compared to electron measurements in the range 49 to 81 keV from the Wind 3DP SST instrument for the 20 October 2002 14:30 solar energetic particle event \citep[adapted from][]{droge_2018}. }
\end{figure}

The pitch-angle distribution of these young electrons is plotted in  Fig. \ref{fig:pitch_angle_dist_D} and is compared with measurements from  the Wind spacecraft during a 9-minute interval following the peak of the 20 October 2002 solar event. In order to temper the dependence on the initial condition (in our case, $v_{\perp}=0$ and therefore $\mu_{\rm B} = 0$), only electrons that experienced at least one collision were retained in Fig. \ref{fig:pitch_angle_dist_D}. This is the reason for the deficit of particles for $\alpha \rightarrow 180^\circ$. Other parts of the distribution are likely slightly affected by choice of the initial condition, as the mean free path $\lambda_\parallel=0.5\:{\rm AU}$ is not much shorter than the distance between the injection point and the observation point at 1 AU.
It is interesting to note that the differences with respect to the measurements of \cite{droge_2018} are  minor, despite the relative arbitrariness of the wind model, initial conditions, mean free path, and injection point adopted in the model. 
\begin{figure}[h]
\includegraphics[width=0.43\textwidth]{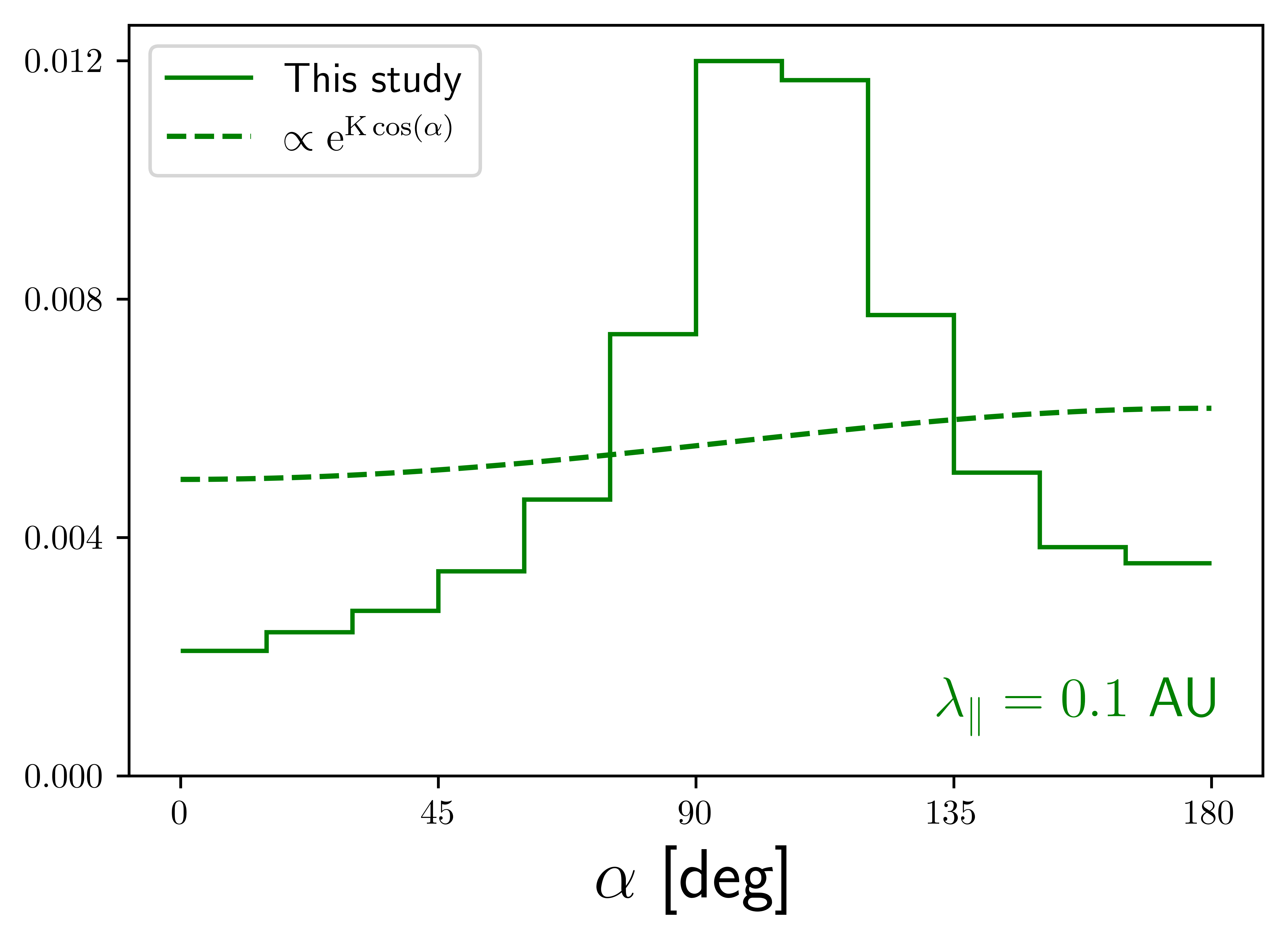}
\includegraphics[width=0.43\textwidth]{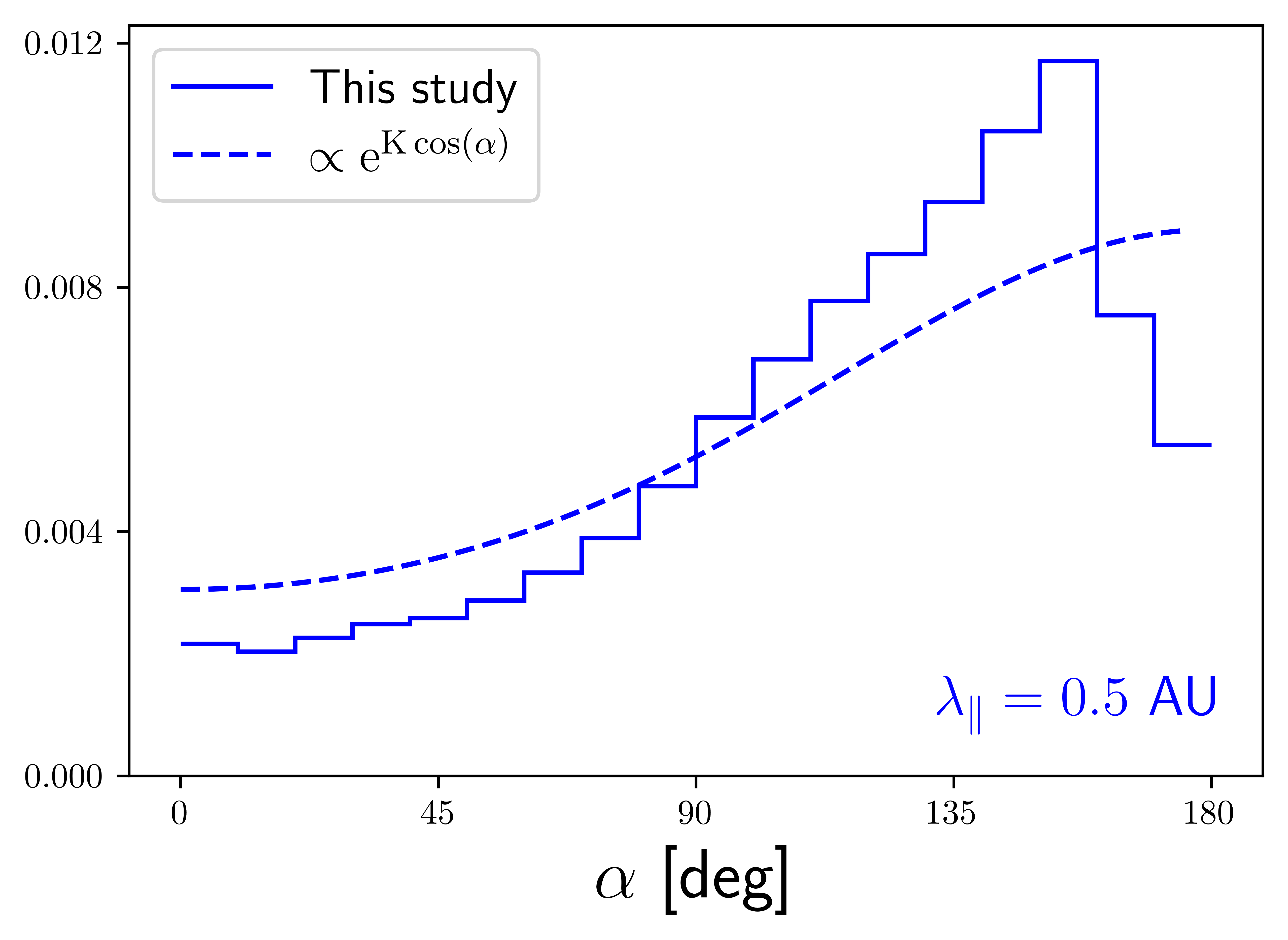}
\includegraphics[width=0.43\textwidth]{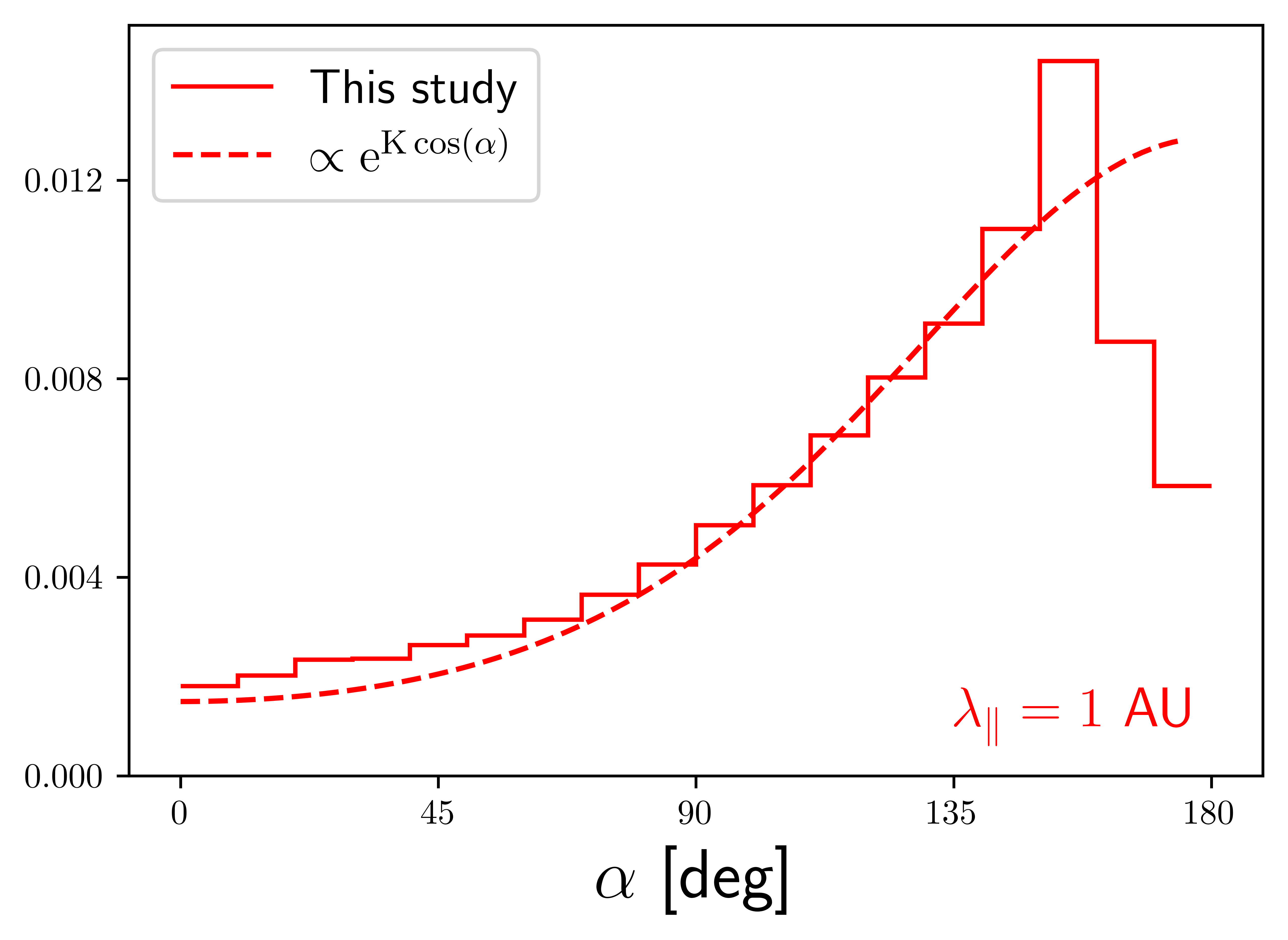}
 \caption{\label{fig:pitch_angle_dist_Z} 
 Normalized steady-state (i.e., all ages) pitch-angle distribution for 81 keV electrons with $\lambda_{\|} = 0.1 \:{\rm AU} $ (top), $\lambda_{\|} = 0.5 \:{\rm AU} $ (middle), and $\lambda_{\|} = 1 \:{\rm AU} $ (bottom). The constant Knudsen predictions from a simple Fokker-Planck model are plotted as well \citep{Zaslavsky_etal_2024}, with the Knudsen numbers K=0.11 (top), K=0.54 (middle), and K=1.08 (bottom) computed from the model values at 1AU (see Table \ref{tab:swprop}). }
\end{figure}
The angular distribution of young electrons is more strongly field aligned than the steady-state distribution, as expected, including electrons of all ages (see the inset in  Fig. \ref{fig:pitch_angle_dist_D} for $\lambda_{\|}=0.5$ AU). The latter is also shown in the middle panel of Fig. \ref{fig:pitch_angle_dist_Z}, where it is compared to the distribution ${\rm K}\exp({\rm K}\cos\alpha)/(2\sinh({\rm K}))$ obtained by \cite{Zaslavsky_etal_2024} from a Fokker-Planck-type equation  in the limit of a spatially constant Knudsen number ${\rm K}=\lambda_{\|}/L_{\rm B}$. The top and bottom panels of Fig. \ref{fig:pitch_angle_dist_Z} display the results for the cases $\lambda_{\|}=0.1$ AU and $\lambda_{\|}=1$ AU, respectively.
We note that as $\lambda_{\|}$ decreases, the distribution becomes isotropic in the Fokker-Planck model, but not in our particle model. As already discussed thoroughly at the end of Sect. \ref{subsec:pda}, this is a property of the scattering law (\ref{eq_alpha_iso}) when it is applied to particles that aer constrained to move in one dimension.

\section{\label{sec:conclusion}Conclusion}
The trajectories and the distribution of energetic test electrons propagating in the field of a simulated solar wind were obtained by integrating the relativistic GCA equations of motion using the $\mathcal{O}(\Delta t^3)$ predictor-corrector scheme that was first proposed by \cite{Mignone_etal_2023}. In order to mimic their interaction with small-scale plasma turbulence, which is not included in the solar wind simulation, particles were also scattered by hard-sphere type collisions. 
Following the consensus on the scattering mean free path of high-energy particles in the solar wind \citep{Palmer_1982,Bieber_1994}, we adopted a mean free path $\lambda_\parallel$ of the order of the astronomical unit.  

In our case study, we computed the evolution of $81\:{\rm keV}$ test electrons continuously injected at r=0.28 AU and $24^{\circ}$ latitude on a corotating magnetic field line in a MHD simulation of a standard steady solar wind. For a parallel mean free path $\lambda_\parallel = 0.5\:{\rm AU}$, the flux of particles observed near 1AU is roughly an exponential function of age, the flux of particles aged 30h is lower by some $10^4$  than the flux of particles aged 1h (see the top of Fig. \ref{fig:ek_and_force_4}). The main outcomes are summarized below.
\begin{enumerate}
    \item  The pitch-angle distribution of young electrons (younger than 20 min and representative of the electrons observed minutes after a solar particle event) compares well with the distribution of electrons in the energy range 49 to 81 keV observed by the WIND spacecraft after the 20 October 2002 14:30 SEP events. This confirms that $\lambda_{\|}=0.5$ AU is a good estimate of the mean free path of $\mathcal{O}(10^2\:{\rm keV})$ electrons near 1AU (see, e.g., \cite{Palmer_1982,Bieber_1994}). 
    \item Electrons systematically lose energy due to their slow poleward motion against the advection electric field because of the curvature drift and the gradient drift.   
    \item For electrons younger than 4h, the drift-induced loss rate is a relatively fast rate of 0.7 keV/h.
    \item For electrons older than 4h, the drift-induced loss rate is lower by a factor 2.
    \item Adiabatic cooling is dominant for particles younger than 1.5 h, for which it is about 1.8 keV/h given a total (drift + adiabatic) loss  rate of 2.1 keV/h. 
    \item  For particles aged 1.5 h, drift and adiabatic loss rates are comparable.  
    For particles older than 1.5 h, the adiabatic loss rate is of secondary importance as its contribution decreases with age roughly as $v_{\rm sw} / \sqrt{v t_{\rm a} \lambda_{\|}}$
\end{enumerate}

Our test-particle propagation method differs from the method introduced by \cite{Marsh_etal_2013} that was subsequently used by \cite{Dalla_etal_2013, Dalla_etal_2015}. In these works, particles were advanced in the electromagnetic field of an analytic wind model by integrating the general form of the equation of motion. Instead, we propagated particles in the electromagnetic field of a simulated solar wind, with the possibility to include dynamical effects such as CMEs. Instead of integrating the full equations of motion, we also integrated the equations in the guiding-center approximation, which allowed us to remove the extremely restrictive requirement of having to use a time step much shorter than the gyration period. This allowed us to use time steps that were longer by several orders and resulted in substantially faster simulations and smaller accumulated numerical errors.   

This work is a first step of a work in progress. Taking advantage of the unprecedentedly large number of  spacecraft that currently travel through the inner heliosphere, our next step is to apply our method to the case of relativistic electrons or protons in the field of a nonstationary solar wind that is traversed by one or several CMEs.

\begin{acknowledgements}
The authors wish to thank Arnaud Zaslavsky for fruitful discussions. 
Ahmed Houeibib is supported by the CNES (Centre National d'Études Spatiales).  This work has been financially supported by the PLAS@PAR project and by the National Institute of Sciences of the Universe (INSU). 
\end{acknowledgements}

\bibliographystyle{aa}
\bibliography{aa}

\end{document}